\documentclass[english,aps,prl,superscriptaddress,twocolumn]{revtex4-1}
\usepackage{mathptmx}
\usepackage[T1]{fontenc}
\usepackage[latin9]{inputenc}
\usepackage{verbatim}
\usepackage{amsmath}
\usepackage{graphicx}
\usepackage{amssymb}

\makeatletter
\@ifundefined{textcolor}{}
{%
 \definecolor{BLACK}{gray}{0}
 \definecolor{WHITE}{gray}{1}
 \definecolor{RED}{rgb}{1,0,0}
 \definecolor{GREEN}{rgb}{0,1,0}
 \definecolor{BLUE}{rgb}{0,0,1}
 \definecolor{CYAN}{cmyk}{1,0,0,0}
 \definecolor{MAGENTA}{cmyk}{0,1,0,0}
 \definecolor{YELLOW}{cmyk}{0,0,1,0}
 }

\makeatother

\usepackage{babel}

\begin{document}

\title{Transition absorption as a mechanism of surface photoelectron emission
from metals}

\author{Alexander V. Uskov}

\email{alexusk@lebedev.ru}

\affiliation{P. N. Lebedev Physical Institute, Russian Academy of Sciences, Leninskiy
Pr. 53, 119333 Moscow, Russia }

\affiliation{Advanced Energy Technologies Ltd, Skolkovo, Novaya Ul. 100, 143025
Moscow Region, Russia }

\author{Igor V. Smetanin}

\affiliation{P. N. Lebedev Physical Institute, Russian Academy of Sciences, Leninskiy
Pr. 53, 119333 Moscow, Russia }

\author{Igor E. Protsenko}

\affiliation{P. N. Lebedev Physical Institute, Russian Academy of Sciences, Leninskiy
Pr. 53, 119333 Moscow, Russia }

\affiliation{Advanced Energy Technologies Ltd, Skolkovo, Novaya Ul. 100, 143025
Moscow Region, Russia }

\affiliation{National Research Nuclear University MEPhI, Kashirskoe highway, 31,
115409 Moscow, Russia }

\author{Renat Sh.~Ikhsanov}

\affiliation{Research Institute of Scientific Instruments (RISI), State Atomic
Energy Corporation ROSATOM, Lytarkino, Turaevo, str. 8, 140080 Moscow
Region, Russia }

\author{Sergei V. Zhukovsky}

\email{sezh@fotonik.dtu.dk}

\affiliation{DTU Fotonik -- Department of Photonics Engineering, Technical University
of Denmark, {\O}rsteds Plads 343, DK-2800 Kgs. Lyngby, Denmark}

\affiliation{ITMO University, Kronverksky pr. 49, St. Petersburg, 197101, Russia}

\author{Viktoriia E. Babicheva}

\affiliation{ITMO University, Kronverksky pr. 49, St. Petersburg, 197101, Russia}

\affiliation{Birck Nanotechnology Center, Purdue University, 1205 West State Street,
West Lafayette, IN, 47907-2057 USA}
\begin{abstract}
\emph{Transition absorption} of electromagnetic field energy by an
electron passing through a boundary between two media with different
dielectric permittivities is considered both classically and quantum
mechanically. It is shown that transition absorption can make a substantial
contribution to the process of electron photoemission from metals
due to the surface photoelectric effect.
\end{abstract}
\maketitle
Transition \emph{emission}, first theoretically predicted in the work
of Ginzburg and Frank in 1946 \cite{1-GinzJETP}, takes place when
an electron passes through a boundary between two media with different
dielectric permittivities. Transition emission was studied in a large
number of papers and was applied, e.g., for charged particle detection
in high energy physics \cite{2-Book}. It is obvious than the inverse
process of transition \emph{absorption} should also exist, i.e., an
electron passing through a boundary between two media with different
permittivities should be able to absorb the energy of electromagnetic
field if such field exists at the boundary. Transition absorption
is particularly interesting in relation to the problem of electron
photoemission from metallic nanoparticles or nanoantennas \cite{3-Nordl11,4-UskovUFN,5-ourNanoscale,6-Govorov,7-Nordl14,8-ourPlas,9-ourPRX}.
One of the photoemission mechanisms (the surface photoelectric effect)
was found to depend both on the potential step and on the dielectric
permittivity step at the interface between a metal and another medium
(dielectric, semiconductor, or vacuum) \cite{4-UskovUFN,5-ourNanoscale}.
The dependence on the permittivity step suggests that transition absorption
can play an important role in photoelectric processes and should be
carefully investigated. 

In this Letter, we perform a qualitative analysis of transition absorption
and derive simple expressions that characterize this effect both classically
and quantum mechanically. We start by considering the electromagnetic
field energy absorption by a classically described electron passing
through a boundary between two media, following a method used earlier
for the analysis of the anomalous skin effect \cite{10-GinzUFN,11-Book}.
Then we calculate the quantum mechanical probability for such an electron
to absorb a photon from the electromagnetic field and show that the
quantum mechanical results for the absorbed energy converge to the
classical ones in the limit $\hbar\to0$. Finally, we identify the
role of transition absorption in the surface photoelectric effect
at a metal boundary, and demonstrate that this role is important.

We note that Brodsky and Gurevich in Refs.~\cite{12-BrodskyJETP,13-Book}
presented general formulas for the probability amplitudes of electron
photoemission, which included the roles of both potential step and
dielectric permittivity step, as well as accounted for the change
of effective electron mass at the metal boundary. However, they did
not perform any specific study of the permittivity step influence
and only used their general formulas to arrive at more specific expressions,
which only took into account the influence of the potential step.
Therefore Refs.~\cite{12-BrodskyJETP,13-Book} did not touch upon
the subject of transition absorption, nor upon its role in photoemission.

\begin{figure}
\includegraphics[width=0.5\columnwidth]{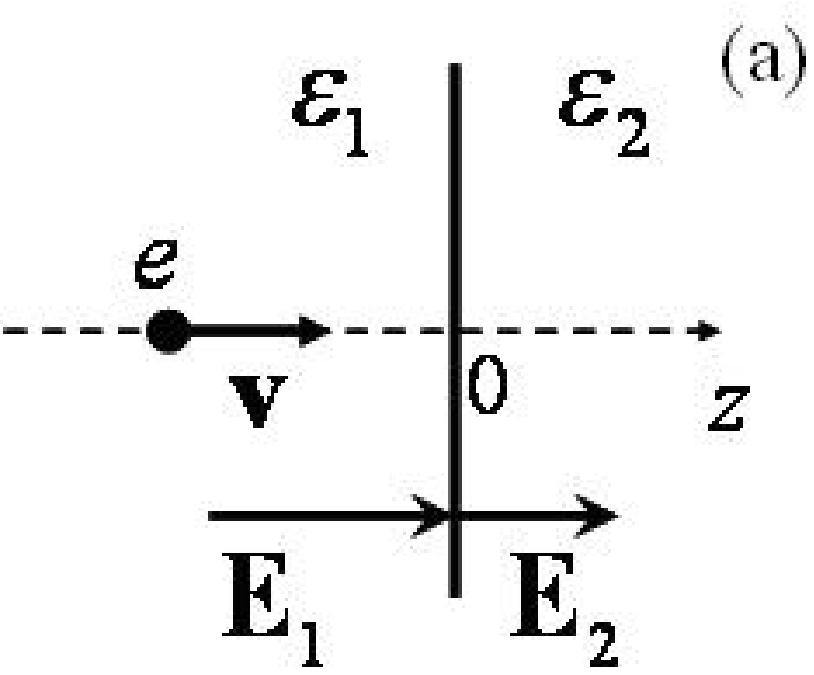}\includegraphics[width=0.5\columnwidth]{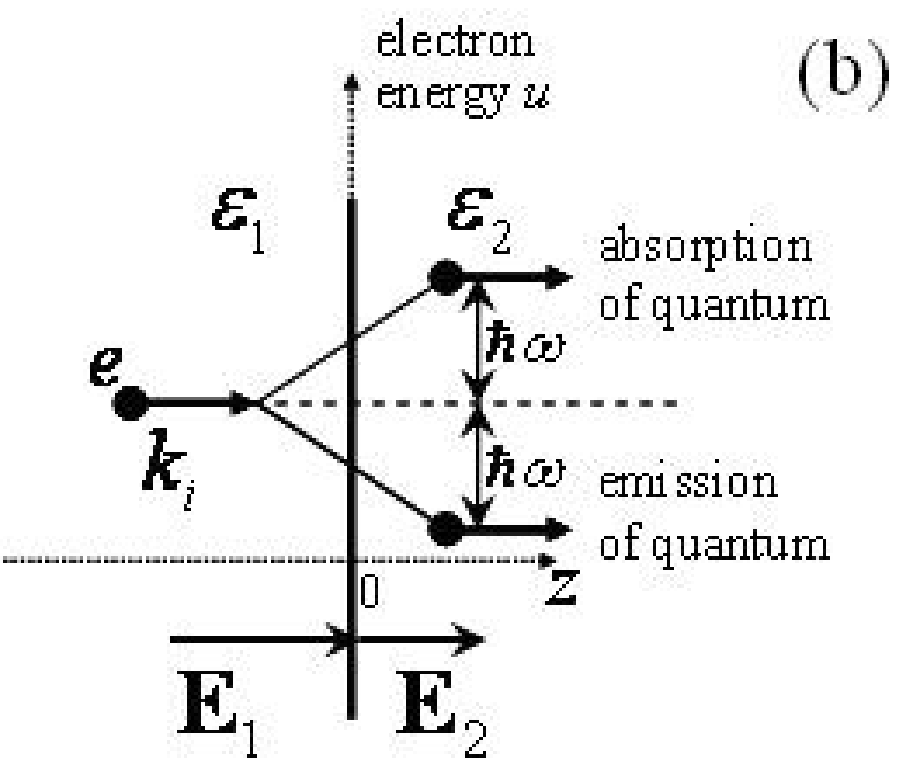}

\caption{(a) An electron (a classical charged particle) moves along a straight
line normal to the boundary between two media with different permittivities
and passes through that boundary. (b) The energy diagram of a corresponding
quantum mechanical problem: an electron impinging upon a boundary
between two media can either absorb or emit a quantum $\hbar\omega$
as it passes through the boundary. \label{FIG:1}}
\end{figure}

Consider first the classical description of transition absorption,
following the model in \cite{6-Govorov,7-Nordl14}. We suppose that
the half-space $z<0$ is filled with medium 1 with dielectric permittivity
$\varepsilon_{1}$ and the half-space $z>0$ is filled with medium
2 with permittivity $\varepsilon_{2}$, as shown in Fig.~\ref{FIG:1}(a).
We further assume that electric field $E_{1}$ and $E_{2}$ exists
in the medium 1 and 2, respectively, and that the field is time-harmonic
with frequency $\omega,$ uniform in the $x$--$y$ plane, and polarized
along the $z$-axis in both media: \begin{equation}
\mathbf{E}_{1,2}(t,\mathbf{r})=\hat{\mathbf{z}}E_{1,2}(t,z)=\hat{\mathbf{z}}E_{1,2}(z)e^{-i\omega t}+\text{c.c.}\label{eq:1}\end{equation}
The complex field amplitudes $E_{1}$ and $E_{2}$ satisfy the boundary
condition at $z=0$:\begin{equation}
\varepsilon_{1}E_{1}=\varepsilon_{2}E_{2}.\label{eq:2}\end{equation}
It is obvious that the field in both media can be represented as \begin{equation}
E\left(t,z\right)=\left[E_{1}(z)+\left(E_{2}(z)-E_{1}(z)\right)\cdot\Theta\left(z\right)\right]e^{-i\omega t}+\text{c.c.},\label{eq:3}\end{equation}
where $\Theta(z)$ is the Heaviside unit step function. 

Suppose that the electron is moving in the field given by Eq.~\eqref{eq:3}
from $z=-\infty$ to $z=\infty$, passing the interface between the
media at $z=0$. The electron motion is described by the equations
\begin{equation}
\begin{gathered}\dot{v}=eE\left[t,z(t)\right]m^{-1},\\
\dot{z}=v(t),\end{gathered}
\label{eq:4}\end{equation}
where $E[t,z(t)]$ is the field acting on the electron located at
$z(t)$ at time $t$, $m$ is the electron mass, and $e=-|e|$ is
its charge. At $t=-\infty$ (or $z=-\infty$) the electron moves at
the constant initial velocity $v_{0}$. We assume that the field is
adiabatically switched on and off along the $z$-axis, i.e., $E_{1}(z)=E_{1}e^{\kappa_{1}z}$
for $z<0$ and $E_{2}(z)=E_{2}e^{-\kappa_{2}z}$ for $z>0$, applying
the limits $\kappa_{1,2}\to0$ at the final step. For sufficiently
weak fields in Eq.~\eqref{eq:3} so that the electron velocity is
only weakly affected, the solution of Eqs.~\eqref{eq:4} can be found
using the perturbation theory by expanding the variables $z(t)$ and
$v(t)$ in terms of the weak field amplitudes: \begin{equation}
\begin{gathered}v(t)=v_{0}+v_{1}(t)+v_{2}(t)+\ldots,\\
z(t)=z_{0}(t)+z_{1}(t)+z_{2}(t)+\ldots,\end{gathered}
\label{eq:5}\end{equation}
where\begin{equation}
z(t),v(t)\propto|E_{1,2}|^{n},\quad n=0,1,2,\ldots\label{eq:6}\end{equation}
In the zero-order approximation we get \begin{equation}
z_{0}(t)=v_{0}(t-t_{s}),\label{eq:7}\end{equation}
where $t_{s}$ is the moment of time when the electron would pass
the boundary if no field were present. Without loss of generality
we can assume that $t_{s}=0$ \cite{6-Govorov,7-Nordl14}.

One can find that the difference between the electron's energy in
the two media (the energy change as the electron passes through the
boundary) is non-vanishing starting from the second order of the perturbation
series and has the form \begin{equation}
\begin{gathered}\Delta W_{\text{classic}}=\left\langle mv^{2}(+\infty)/2-mv_{0}^{2}(+\infty)/2\right\rangle \\
\approx(m/2)\left\langle v_{1}^{2}(+\infty)\right\rangle +mv_{0}\left\langle v_{2}(+\infty)\right\rangle ,\end{gathered}
\label{eq:8}\end{equation}
where $\langle\ldots\rangle$ denotes the averaging with respect to
the phase $\phi_{1}$ of the complex amplitude $E_{1}=|E_{1}|e^{-i\phi_{1}}$
\cite{6-Govorov,7-Nordl14}.

Substituting the expansions in Eqs.~\eqref{eq:5} into Eqs.~\eqref{eq:4},
in the first order of the perturbation theory we find \begin{equation}
v_{\text{1}}\left(+\infty\right)=\frac{ie\left|E_{1}\right|}{m\omega}\left[e^{-i\phi_{1}}\left(1-\frac{\varepsilon_{1}}{\varepsilon_{2}}\right)-e^{+i\phi_{1}}\left(1-\frac{\varepsilon_{1}^{*}}{\varepsilon_{2}^{*}}\right)\right],\label{eq:9}\end{equation}
and therefore,\begin{equation}
\left\langle v_{1}^{2}\left(+\infty\right)\right\rangle =\frac{2e^{2}}{m^{2}\omega^{2}}\left|E_{1}\right|^{2}\left|\frac{\varepsilon_{1}}{\varepsilon_{2}}-1\right|^{2}.\label{eq:10}\end{equation}
In the second order the expression for $v_{2}(+\infty)$ turns out
to be rather bulky, so we only give the explicit form of its average:
\begin{equation}
\left\langle v_{2}\left(+\infty\right)\right\rangle =\frac{e^{2}}{m^{2}\omega^{2}v_{0}}\left|E_{1}\right|^{2}\left|\frac{\varepsilon_{1}}{\varepsilon_{2}}-1\right|^{2}.\label{eq:11}\end{equation}
 Substituting Eqs.~\eqref{eq:10} and \eqref{eq:11} into Eq.~\eqref{eq:8},
we finally get \begin{equation}
\Delta W_{\text{classic}}=\frac{2e^{2}}{m\omega^{2}}\left|E_{1}\right|^{2}\left|\frac{\varepsilon_{1}}{\varepsilon_{2}}-1\right|^{2}=\frac{2e^{2}}{m\omega^{2}}\left|E_{2}-E_{1}\right|^{2}.\label{eq:12}\end{equation}

Equation \eqref{eq:12} shows that the average energy absorbed by
the electron passing from medium 1 to medium 2 is proportional to
the square of the dielectric permittivity difference, i.e., $\Delta W\propto|\varepsilon_{1}-\varepsilon_{2}|^{2}$.
This constitutes the effect of transition absorption. It exists irrespective
of the sign of $\varepsilon_{1}-\varepsilon_{2}$, i.e., irrespective
of whether the electron passes from medium 1 to medium 2 or the other
way around. As will be shown below, the classical expression \eqref{eq:12}
remains valid in the quantum case if the quantum energy is much less
than the electron energy ($\hbar\omega\ll u_{i}$). 

Note that the first-order correction to the electron velocity is sufficient
for the description of the anomalous skin effect \cite{12-BrodskyJETP,13-Book}
because it is typically considered for the normal incidence of light
on the metal boundary, so light causes the electrons to oscillate
parallel to the boundary. In such a case, at least for electrons normally
incident on the boundary and experiencing specular reflection from
it, the expression for $\Delta W$ similar to Eq.~\eqref{eq:8} contains
no term proportional to $v_{2}(+\infty)$, since the initial velocity
$\mathbf{v}_{0}$ and the velocity $\mathbf{v}_{2}(+\infty)$ are
perpendicular to each other. On the contrary, for the transition absorption
in the geometry shown in Fig.~\ref{FIG:1}(a) the term proportional
to $\langle v_{2}(+\infty)\rangle$ in Eq.~\eqref{eq:8} is non-vanishing
and equal to the term proportional to $\langle v_{1}^{2}(+\infty)\rangle$.
The substantial contribution of both these terms requires the application
of the regular perturbation theory up to the second order. 

Now let us consider the transition absorption effect quantum mechanically,
showing that quantum mechanical calculations also converge to Eq.~\eqref{eq:12}.
Figure \ref{FIG:1}(b) shows the set-up of the problem. An electron
with energy $u_{i}=\hbar^{2}k_{i}^{2}/2m$ ($k_{i}$ being the electron
wave number) is normally incident on the boundary between two media.
When passing through the boundary, the electron can either absorb
or emit a quantum of electromagnetic energy $\hbar\omega$ with probabilities
$p_{+}$ and $p_{-}$, respectively. The energy of the electron then
becomes $u_{\pm}=\hbar^{2}k_{\pm}^{2}/2m=u_{i}\pm\hbar\omega$. The
probabilities $p_{\pm}$ can be found using the Fermi golden rule
as \begin{equation}
p_{\pm}=p_{\pm}\left(k_{o}\right)=\frac{m}{\hbar^{3}}\cdot\frac{1}{k_{\pm}}\cdot\left|\left\langle i\right|H'_{\pm}\left|f_{\pm}\right\rangle \right|^{2},\label{eq:13}\end{equation}
where the initial and final wave functions of the electron are, \begin{equation}
\left|i\right\rangle =v_{i}^{-1/2}e^{+ik_{i}z},\quad\left|f_{\pm}\right\rangle =e^{+ik_{\pm}z},\label{eq:14}\end{equation}
the subscripts $_{+}$ and $_{-}$ corresponding to photon absorption
and emission, respectively. The interaction Hamiltonians $H'_{\pm}$
are \begin{equation}
H'_{+}=\frac{i\hbar e}{2m}\frac{\partial A_{z}\left(z\right)}{\partial z}+\frac{i\hbar e}{m}A_{z}\left(z\right)\frac{\partial}{\partial z},\quad H'_{-}=\left(H'_{+}\right)^{*}.\label{eq:15}\end{equation}
Here the vector potential component of the electromagnetic field $A_{z}(z)$
is equal to $A_{1}$ for $z<0$ and to $A_{2}$ for $z>0$; this potential
is related to the electric field $E_{z}(z)$ as $A_{z}(z)=-iE_{z}(z)/\omega$.
Using Eqs.~\eqref{eq:14}--\eqref{eq:15}, Eq.~\eqref{eq:13} results
in \begin{equation}
p_{\pm}=\frac{e^{2}u_{i}}{2m\hbar^{2}\omega^{4}}\cdot\frac{\left(1+\sqrt{1\pm\frac{\hbar\omega}{u_{i}}}\pm\frac{\hbar\omega}{2u_{i}}\right)^{2}}{\sqrt{1\pm\frac{\hbar\omega}{u_{i}}}}\cdot\left|E_{2}-E_{1}\right|^{2}.\label{eq:16}\end{equation}
Similarly to the classical analysis, we assume adiabatically smooth
variation of the field (on-off switching) along the $z$-axis during
the integration over $z$ needed to evaluate the matrix element in
Eq.~\eqref{eq:13}.

Now we can use Eq. \eqref{eq:16} to derive the average energy transferred
from the field to the electron passing through the boundary between
two media: \begin{equation}
\Delta W_{\text{quantum}}=\hbar\omega\cdot\left[p_{+}-p_{-}\right].\label{eq:17}\end{equation}
 In the classical limit $\hbar\omega/u_{i}\ll1$ Eqs.~\eqref{eq:16}
and \eqref{eq:17} result in \begin{equation}
\Delta W_{\text{quantum}}\approx2\frac{e^{2}}{m\omega^{2}}\left|E_{2}-E_{1}\right|^{2},\label{eq:18}\end{equation}
which coincides with the classical result given by Eq.~\eqref{eq:12}. 

\begin{figure}
\includegraphics[width=0.5\columnwidth]{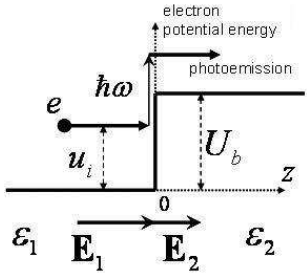}

\caption{Photoelectron emission from metal: an electron with energy $u_{i}$
absorbs a photon with energy $\hbar\omega$ during collision with
the potential barrier\textbf{ $U_{b}$} at the metal boundary, overcomes
the barrier and leaves the metal. \label{FIG:2}}
\end{figure}

Now let us show how transition absorption influences the electron
photoemission from metals. In the surface photoelectric effect scenario
investigated, e.g., in \cite{4-UskovUFN,5-ourNanoscale,12-BrodskyJETP,13-Book},
an electron with the wave number $k_{i}$ and energy $u_{i}=\hbar^{2}k_{i}^{2}/2m$
is normally impinging on a potential step with height $U_{b}$, which
models a boundary between a metal and some adjacent medium (semiconductor,
dielectric or vacuum), into which electrons can be emitted from metal,
as seen in Fig.~\ref{FIG:2}. The structure contains a light wave
with electric field amplitudes $E_{1}$ and $E_{2}$ in medium 1 (metal)
and medium 2 (e.g. semiconductor), respectively. As an electron collides
with the potential step, it can absorb a photon with energy $\hbar\omega$.
If the electron's resulting energy $u_{i}+\hbar\omega$ exceeds $U_{b}$,
then the electron can leave the metal. Following, e.g., the steady-state
perturbation theory \cite{4-UskovUFN}, one can derive the probability
of photoelectron emission \begin{equation}
\begin{gathered}p_{\text{PE}}=0.25K_{\text{PE}}\sqrt{u_{i}+\hbar\omega-U_{b}}\\
\times\left|\left(E_{2}+E_{1}\right)U_{b}-\left(E_{2}-E_{1}\right)\left(\sqrt{u_{i}+\hbar\omega}+i\sqrt{U_{b}-u_{i}}\right)^{2}\right|^{2},\end{gathered}
\label{eq:19}\end{equation}
where \begin{equation}
K_{\text{PE}}=\frac{8e^{2}}{m\hbar^{2}\omega^{4}U_{b}^{2}}\cdot\frac{\sqrt{u_{i}}\cdot\left|\sqrt{u_{i}}-i\sqrt{U_{b}-u_{i}}\right|^{2}}{\left|\sqrt{u_{i}+\hbar\omega}+\sqrt{u_{i}+\hbar\omega-U_{b}}\right|^{2}}.\label{eq:20}\end{equation}

The first term inside the modulus on the second line of Eq.~\eqref{eq:19},
$\left(E_{2}+E_{1}\right)U_{b}$, describes the contribution to photoemission
stemming from the process of electron collision with the potential
step, the so-called inverse bremsstrahlung \cite{14-BookFed}. This
term disappears for $U_{b}\to0$ and contains the sum of the field
amplitudes, $E_{2}+E_{1}$, i.e., the inverse bremsstrahlung persists
if the two media have equal permittivities ($\varepsilon_{1}=\varepsilon_{2}$).
On the contrary, the remaining second term in Eq.~\eqref{eq:19}
is proportional to $E_{2}-E_{1}$, and thus vanishes if $\varepsilon_{1}=\varepsilon_{2}$.
It is this term that corresponds to the contribution of the transition
absorption to photoemission; one can see that it converges to $p_{+}$
in Eq.~\eqref{eq:16} in the limit $U_{b}\to0$. One can also see
from Eq.~\eqref{eq:19} that the contributions from these two terms
are not additive but undergo quantum interference of their complex
probability amplitudes. %
{}

\begin{figure}
\includegraphics[bb=2445bp 10bp 2756bp 250bp,clip,width=1\columnwidth]{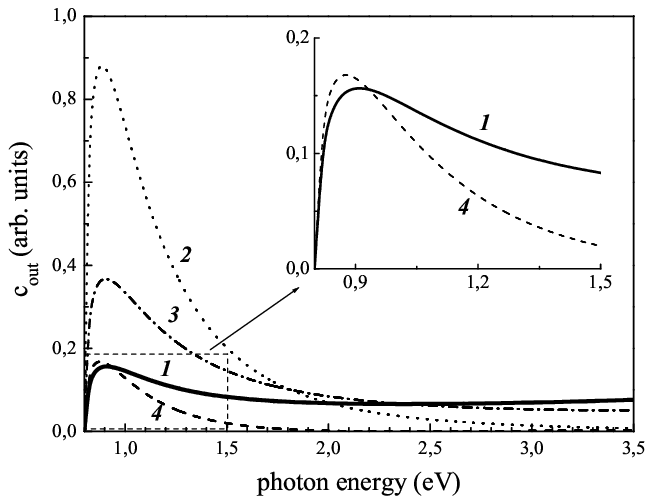}

\caption{Spectrum of the photoemission coefficient $c_{\text{out}}$ defined
in Eq.~\eqref{eq:21} for Au/GaAs interface with $u_{i}=5.5\,\text{eV}$
and $U_{b}=u_{i}+0.8\,\text{eV}$. Full contribution taking into account
both the inverse bremsstrahlung term and the transition absorption
term in Eq.~\eqref{eq:19} (curve \emph{1}, solid line) is compared
with the result of keeping only the inverse bremsstrahlung term (curve
\emph{4}, dashed line) and only the transition absorption term (curve
\emph{3}, chain line); the dotted line (curve \emph{2}) shows the
special case of $E_{1}=E_{2}$ (neglecting the dielectric permittivity
step at the boundary). The inset shows the behavior of curves \emph{1}
and \emph{4} near the photoemission threshold. \label{FIG:3}}
\end{figure}

Consider an example when photoelectrons are emitted from a metal into
a semiconductor \cite{8-ourPlas,9-ourPRX}. The incident electron
energy is taken to be $u_{i}=5.5\,\text{eV}$ (the typical Fermi energy
in gold and silver), and the potential step is $U_{b}=u_{i}+0.8\,\text{eV}$
(meaning that $p_{\text{PE}}>0$ only for $\hbar\omega>0.8\,\text{eV}$).
The dielectric permittivity of the semiconductor is $\varepsilon_{2}=13$
(GaAs), whereas for the metal the Drude model is assumed so that $\varepsilon_{1}=1-(\hbar\omega_{p})^{2}/[(\hbar\omega)^{2}+i(\hbar\gamma_{p})(\hbar\omega)]$
with $\hbar\omega_{p}=8.9\,\text{eV}$ and $\hbar\gamma_{p}=0.07\,\text{eV}$,
which are typical values for gold.

The calculation results are presented in Fig.~\ref{FIG:3} in the
form of the spectral dependence of the surface photoemission coefficient
defined as\begin{equation}
c_{\text{out}}(\hbar\omega,U_{b},u_{i},E_{1}/E_{2})=p_{\text{PE}}/|E_{2}|^{2}.\label{eq:21}\end{equation}
This coefficient is convenient because it does not depend on the field
strength (unlike the probability $p_{\text{PE}}$) and thus characterizes
only the media properties as well as the electron and photon energies.
The solid line in Fig.~\ref{FIG:3} shows the full $c_{\text{out}}$
calculated according to Eqs.~\eqref{eq:19}--\eqref{eq:20}, taking
both effects (inverse bremsstrahlung and transition absorption) into
account. The dashed line shows the case when only the inverse bremsstrahlung
is retained, i.e., only the first term is kept inside the modulus
in Eq.~\eqref{eq:19}. The chain line similarly shows the case when
only the transition absorption {[}the second term in Eq.~\eqref{eq:19}{]}
is retained.

The comparison between these three curves shows that transition absorption
contribution to the overall photoemission is generally the strongest.
Interestingly, the account for transition absorption causes the total
photoemission coefficient to decrease close to the threshold value
of $\hbar\omega$ (slightly over 0.8 eV, see inset in Fig.~\ref{FIG:3}).
This results from destructive interference of probability amplitudes
in Eq.~\eqref{eq:19} caused by the minus sign between the two terms.
As $\hbar\omega$ increases further away from the threshold, the inverse
bremsstrahlung contribution becomes vanishingly small. Indeed, the
Drude formula predicts that the absolute value of $\mathrm{Re}\,\varepsilon_{1}$
decreases with the increase of $\hbar\omega$, and that $\mathrm{Re}\,\varepsilon_{1}$
remains negative. As a result, the first term inside the modulus in
Eq.~\eqref{eq:19} has a minimum when \begin{equation}
\mathrm{Re}\,(\varepsilon_{1})+\varepsilon_{2}=0,\label{eq:22}\end{equation}
which takes place at the values of $\hbar\omega$ around 2.5 eV for
the chosen parameters. Nevertheless, the full photoemission coefficient
is non-zero in that range, decreasing to an almost constant value
after the peak at 0.9 eV instead. Such behavior is the direct consequence
of transition absorption. Therefore we can conclude that transition
absorption makes a dominant contribution to photoelectron emission
throughout the entire considered spectral range for the chosen parameters. 

Note that the vanishing of the inverse bremsstrahlung contribution
given by Eq.~\eqref{eq:22} takes place in the optical range ($\hbar\omega\simeq2\ldots3\,\text{eV}$)
for the internal photoelectric effect when the medium 2 has a relatively
large $\varepsilon_{2}$, as considered here. For the external photoelectric
effect when that medium is vacuum, Eq.~\eqref{eq:22} is satisfied
for photon energies larger by 3--4 times. In such a case the frequency
dependence of all curves in Fig.~\ref{FIG:3} would be dominated
by the coefficient in front of the modulus in Eq.~\eqref{eq:19},
so all the three lines (solid, dashed and chain) would become approximately
congruent to each other.

Finally, the dotted line in Fig.~\ref{FIG:3} shows a frequently
used approximation (see, e.g., \cite{13-Book} and references therein)
assuming that the fields inside the metal $E_{1}$ and outside the
metal $E_{2}$ are equal ($E_{1}=E_{2}$). Comparing this curve to
the accurate result of Eq.~\eqref{eq:19}, we see that this approximation
overestimates the coefficient $c_{\text{out}}$ very substantially
(by a factor of about 4) for the values of $\hbar\omega$ such that
$|\varepsilon_{2}/\varepsilon_{1}|\ll1$, and underestimates $c_{\text{out}}$
for higher photon energies. So, we see that the approximation $E_{1}=E_{2}$
can be regarded as only qualitatively correct.

To summarize, we have addressed the problem of transition absorption
of light wave energy by an electron as it passes through a boundary
between two media, and have solved this problem both classically and
quantum mechanically. In the limit when the photon energy is much
less than the electron energy, the classical and quantum results converge
to each other. We have also shown that transition absorption makes
a substantial contribution to the process of surface photoelectron
emission from metallic nanoparticles.

The authors acknowledge the Russian Foundation for Basic Research
(Grant No. 15-02-03152) for financial support. One of us (S.V.Z.)
acknowledges support from the People Programme (Marie Curie Actions)
of the European Union\textquoteright{}s 7th Framework Programme FP7-PEOPLE-2011-IIF
under REA grant agreement No.~302009 (Project HyPHONE).

\end{document}